# State-resolved electron capture in low-energy $Ar^{2+}$-Ar/$N_2$ collisions*

CUI Shucheng[1, 2], XING Dadi[2], ZHU Xiaolong[2, 3, 4, *], ZHAO Dongmei[2], GUO Dalong[2, 3, 4], GAO Yong[2, 3, 4], ZHANG Shaofeng[2, 3, 4], DONG Chenzhong[1, *], MA Xinwen[2, 3, 4]

1. Key Laboratory of Atomic and Molecular Physics & Functional Materials of Gansu Province, College of Physics and Electronic Engineering, Northwest Normal University, Lanzhou 730070, China
2. Institute of Modern Physics, Chinese Academy of Sciences, Lanzhou 730000, China
3. State Key Laboratory of Heavy Ion Science and Technology, Institute of Modern Physics, Chinese Academy of Sciences, Lanzhou 730000, China
4. University of Chinese Academy of Sciences, Beijing 100049, China

**Abstract**

As a fundamental process in atomic physics, charge exchange relies on quantum state-resolved data that is crucial for various fields such as astrophysics and plasma physics. However, there remains a g in the research on multi-electron target systems. This study aims to investigate the dynamic mechanisms of single/double electron capture in collisions between $Ar^{2+}$ ions and Ar atoms or $N_2$ molecules at an energy of 40 keV, thereby supplementing high-precision experimental data in this field. The experiment is conducted on the electron beam ion source (EBIS) platform at the Institute of Modern Physics, Chinese Academy of Sciences, using the cold target recoil ion momentum spectroscopy (COLTRIMS) technique. An ion beam containing ground-state $Ar^{2+}$ ($3s^2 3p^4\ {}^3P$) and metastable $Ar^{2+}$ ($3s^2 3p^4\ {}^1D,\ {}^1S$) is used as the projectile, colliding with a supersonic Ar/ $N_2$ mixed gas target. Three-dimensional momentum of recoil ions is reconstructed through coincidence measurements of recoil ions and scattered ions, and the $Q$-value and scattering angle distribution are calculated. Theoretical comparisons are performed using the molecular Coulombic over barrier model (MCBM).
The results show that there are similarities in the populations of single-electron captured states between the two systems, but the contribution ratios are different: the $Q$-value spectrum of the $Ar^{2+}$-Ar system contains an additional characteristic peak, which corresponds to the process where the projectile ion captures an electron from the 3s orbital of the target while its own 3s electron is excited to the 3p orbital. In contrast, this characteristic peak is absent in the $Ar^{2+}$ - $N_2$ system due to the easy dissociation of excited $N_2^+$ ions. For double-electron capture, both systems are dominated by capturing electrons to the ground state, but only the $Ar^{2+}$ - $N_2$ system shows a significant contribution from excited state populations. The comparison of scattering angles reveals that the higher the capture state of the product ion, the larger the corresponding scattering angle is and the smaller the impact parameter is. This is presumably because electron interactions become more complex at smaller impact parameters, leading to a higher probability of capturing electrons to high-energy levels. In the double-electron capture of the $Ar^{2+}$ - $N_2$ system, only the ground-state channel is populated at small angles (0 – 1.2 mrad). Additionally, electron capture exhibits dependence on impact parameter: as the angle increases (i.e. the impact parameter decreases), the $Q$-value of the capture reaction decreases, indicating that the reaction tends to be more endothermic.

# 1. Introduction

Charge exchange is a fundamental process in atomic physics. The study of this process is not only of great academic value for deepening the basic scattering theory, but also of great significance in many applied science fields such as astrophysics, plasma physics and ion-induced biological radiation effects[1-3]. Research in these fields relies on high-precision collision cross section data, which is not only the basis for building theoretical models of reaction kinetics, but also the core element for understanding the energy transport process of complex many-body systems. The microscopic mechanism of charge exchange involves the transfer of electrons from neutral atoms or molecules to highly charged ions. Since the first discovery of X-ray emission from Hyakutake B2 comet by ROSAT telescope in 1996 and the successful explanation of this observation by charge exchange between highly charged heavy ions in the solar wind and neutral gas in the comet, the charge exchange process has once again become a hot topic in the fields of experimental and theoretical physics[1,4,5].

In the past decades, abundant experimental data and theoretical results have been accumulated on the charge exchange process between helium atom and hydrogen molecule[6-11]. However, the research on multi-electron target system is still insufficient at the experimental and theoretical levels. The main reason is that more target electrons will participate in the collision reaction of multi-electron target, and the collision reaction channel is extremely complex, which leads to the relatively slow progress in theoretical calculation and experimental observation. Monoatomic gas target is an ideal choice for ion and multi-electron target collision experiments because of its simple structure, high purity, controllable parameters and clear theoretical model. Argon was first used as a target gas to study ionization, charge transfer and other basic processes, mainly because argon is the third most abundant inert gas in the atmosphere. In the previous theoretical and experimental studies of argon targets, scholars mainly focused on the collision cross section of ions with argon. For example, Suk et al.[12]studied the collision process of $Ar^{2+}$ ions with Ne, Ar and Kr at 50-200 keV by accelerator technology, and measured the single electron capture cross sections of this process. The experimental results are in good agreement with the theoretical calculation results of Rapp and Francis[13] in the trend of energy variation. The differential cross sections of charge exchange in the interaction of low energy $Ar^{2+}$ ions with He, Ne, Ar were measured by Huber[14] using an energy loss spectrometer. The results show that only a few reaction channels are important, and the total cross sections of charge exchange are strongly dependent on energy. Shields and Moran[15] used time-of-flight technique to measure the electron transfer cross sections for the reactions of low-energy ground state and metastable $Ar^{2+}$ ions with different target gases (Ar, $N_2$, $O_2$, CO, $CO_2$, $CH_4$ and $C_2H_6$), and found that the electron transfer cross sections of individual systems in the ground state and metastable $Ar^{2+}$ reactions were almost the same. However, the energy loss spectrometer is limited to the study of low-energy collisions, and the early time-of-flight technology has low detection efficiency and cannot accurately obtain the quantum state population, which leads to the scarcity of accurate electron capture quantum state selection cross section data in a wider collision energy range. In recent years, great progress has been made in the measurement of absolute charge exchange cross section, with a wider energy range and higher accuracy[16,17], which provides a basis for the study of quantum state-resolved electron capture.

In order to further explore the dynamic mechanism of electron capture process, it is very important to accurately measure the scattering angular distribution of projectile ions after collision. However, the scattering angle produced by the capture reaction of ions with atoms and molecules is usually less than a few milliradians (mrad), which puts forward extremely high requirements for experimental measurement techniques. At the end of the 20th century, the traditional ion momentum spectrometer was improved by German scientists, and the cold target recoil ion momentum spectroscopy (COLTRIMS) was successfully developed by combining with the cryogenic target technology, which was the first time to achieve high-resolution recoil ion momentum measurement[18,19]. Thanks to the development of inverse momentum spectroscopy, the energy and scattering angle of recoil ions can be measured with high resolution at high collision energy.

In terms of theoretical research, the molecular coulombic barrier model (MCBM) is a commonly used method, which can calculate the reaction window and the state population cross section, and qualitatively explain some ion-atom collision phenomena, but it cannot predict the complex quantum effects, the accurate contribution of the weak reaction channel, and the complex collision process in the intermediate energy region[20,21]. A more complex method is the classical trajectory Monte Carlo (CTMC) method, which simulates the collision process by constructing a three-body dynamic system including the incident ion, the target ion, and the target electron, numerically solving Hamilton's equations, and propagating trajectories based on classical mechanics[22]. However, it neglects the detailed molecular state[23] associated with the vibrationally excited levels of the target, and the structure of the incident ion becomes important at energies below 1 keV/amu, resulting in limited application of this method[24]. Another method is the Landau-Zener (LZ) approximation, which can often clarify the general behavior of the cross section and can be used for preliminary estimates of state-selective capture, but its prediction accuracy is limited[25]. At present, the most reliable theoretical framework is the quantum-mechanical close-coupling (CC) method, which can deal with atomic/quasi-molecular structures and their reaction dynamics, but requires extremely stringent computational technology and resources[26].

The contribution of metastable projectile ions is a non-negligible contribution[15,27-30] in the collision test using $Ar^{2+}$ ions as projectile ions, which is also confirmed in the theoretical study[31]. In this study, the single and double electron capture processes of $Ar^{2+}$- Ar/$N_2$ at 40 keV collision energy were systematically studied by using an recoil momentum spectrometer, and the accurate measurement of the scattering angular distribution was realized. In addition, the correlation analysis of quantum state population characteristics is further carried out for the double electron capture process of $Ar^{2+}$-$N_2$ system in the scattering angle range of $0 - 1.2$ mrad.

## 2. Experimental setup

The experiment was carried out on the electron beam ion source (EBIS) experimental platform of the Institute of Modern Physics, Chinese Academy of Sciences[32-35]. The COLTRIMS experimental device on the EBIS platform has been described in detail by Ma et al.[36,37] in related studies, and only a brief description is given in this paper. An EBIS ion source was used to generate and extract $Ar^{2+}$ ion beams, which mainly include three ion states: ground ion $Ar^{2+}(3s^2 3p^4\ ^3P)$, metastable ion $Ar^{2+}(3s^2 3p^4\ ^1D)$, and metastable ion $Ar^{2+}$ $(3s^2 3p^4\ ^1S)$. These ions were accelerated to 40 keV by a high-pressure platform, collimated by a four-jaw collimating slit, and then cross-collided with a supersonic Ar/$N_2$ gas mixture jet. The recoil target ions produced in the collision process were transmitted to a recoil position-sensitive detector (PSD-R) for detection under the guidance of an electric field perpendicular to the direction of the projectile ion beam and the target beam; at the same time, the scattered ions are recorded by an electrostatic analyzer equipped with a position-sensitive detector-projectile (PSD-P). Through the coincidence measurement of recoil ion and scattered ions, the system realizes the event-by-event recording of recoil ion information on the PSD-R detector. Based on the two-dimensional position information of the recoil ion on the PSD-R detector and its arrival time, the three-dimensional momentum distribution of the recoil ion after collision can be reconstructed. For the $Ar^{2+}$-Ar/$N_2$ collision system, the momentum resolution of the experimental setup is about 1 a. u. In this paper, all physical quantities are expressed in atomic units (a. u.).

## 3. Results and Discussion

The longitudinal momentum (along the direction of the projectile ion beam) and the transverse momentum (perpendicular to the direction of the projectile ion beam) of the recoil ion in the final state of the reaction can be extracted from the three-dimensional momentum information of the recoil ion after collision using the vector synthesis method. The analysis of longitudinal momentum can provide the state-selective population information of the projectile ion products; the precise measurement of the transverse momentum can directly obtain the scattering angular distribution, which contains the dynamic characteristics of the populated states in the reaction process. The longitudinal momentum $P_{\text{long}}$ and transverse momentum $P_{\text{trans}}$ of the recoil ion can be determined by the following relations, respectively:

$$P_{\text{long}} = -\frac{Q}{v_{\text{p}}} - \frac{n}{2} v_{\text{p}}, \tag{1}$$

$$P_{\text{trans}} = P_0 \theta, \tag{2}$$

Where $Q$ represents the change in the electron binding energy before and after collision, $v_{\text{p}}$ is the velocity of the projectile ion, $n$ is the number of captured electrons, $P_0$ is the initial momentum of the projectile ion, and $\theta$ is the scattering angle of the projectile ion.

### 3.1 $Ar^{2+}$-Ar collision system

The Fig. 1(a) shows the distribution of the $Q$ values of the single-electron capture process in the $Ar^{2+}$-Ar collision system, where the black scattered points are the experimentally measured data, the red solid line corresponds to the reaction window function calculated by the molecular Coulomb over barrier model (MCBM), and the intersection of the blue vertical line and the reaction window represents the relative cross section size of each capture channel. Three characteristic peak structures (marked as A, B, C) are observed in the spectrum, and the corresponding reaction mechanisms are as follows. Peak A is derived from the capture of the ground state $Ar^{2+}(3s^2p^4\,^3P,^1D,^1S)$ with the ground state Ar ($^1S$): $Ar^{2+}(3s^23p^4\,^3P,^1D,^1S)$ + Ar ($^1S$) → $Ar^+$ ($3s^23p^5\,^2P^0$) + $Ar^+$ ($3s^23p^5\,^2P^0$), corresponding to the $Q$ values of +11.87eV，+13.46 eV and +15.88 eV [38]. The main capture reaction of peak B is $Ar^{2+}$ ($3s^23p^4\,^3P,^1D,^1S$) + Ar ($^1S$) → $Ar^+$ ($3s^3p^6\,^2S$) + $Ar^+$ ($3s^23p^5\,^2P^0$), corresponding to $Q$ values of –1.61 eV, +0.02 eV and +2.40 eV, respectively. In addition, $Ar^{2+}(3s^23p^4\,^3P,^1D,^1S)$ + Ar ($^1S$) → $Ar^+(3s^23p^4nl)$ +$Ar^+(3s^23p^5\,^2P^0)$. The process also made certain contributions. Peak C corresponds to the double excited state capture process: $Ar^{2+}$ ($3s^23p^4\,^3P,^1D,^1S$) + Ar ($^1S$) → $Ar^+$ ($3s3p^6\,^2S$) + $Ar^+$ ($3s3p^6\,^2S$), corresponding to $Q$ values of –15.09 eV, –13.46 eV and –11.08 eV. Like the energy loss spectrometer, in the symmetric collision system, we cannot distinguish whether the electron after the reaction is in the excited state of the recoil ion or the excited state of the scattering ion, that is, we cannot distinguish the cases of $(n = 1, n' > 1)$ and $(n > 1, n' = 1)$, mainly because the $Q$ values of the two are the same.

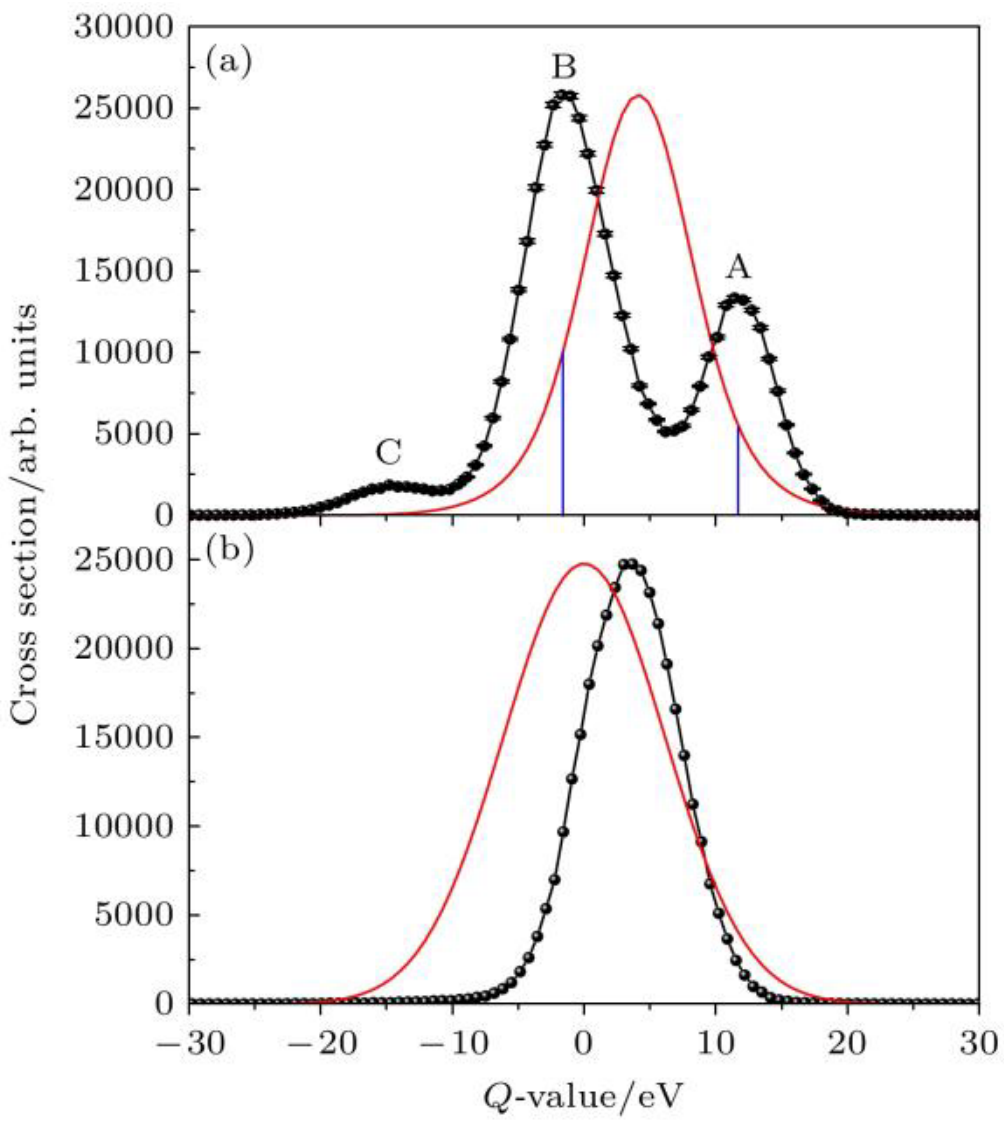

**Figure 1.** $Q$ -value distributions for electron capture processes in $Ar^{2+}$-Ar collisions at 40 keV: (a) Single electron capture; (b) double electron capture.

Based on the theory [21] of MCBM, the single-electron capture dynamics of $Ar^{2+}$-Ar collision system has been systematically analyzed. The reaction window calculated by MCBM can predict the information of state population and relative state selective cross section of ion and atom capture process. The red curve of the Fig. 1 shows the reaction window calculated by MCBM. Since the model is only applicable to the theoretical description of the ground state incident ion, this paper chooses to compare quantitatively with the ground state collision experimental data. The height of the intersection of the blue vertical line and the red solid line in the figure represents the relative size of the reaction cross section. The results of MCBM calculation show that the cross section of electron capture to $Ar^+$ ($3s3p^6\,^2S$) is the largest, followed by $Ar^+$ ($3s^23p^5\,^2P^0$) in $Ar^{2+}$-Ar single-electron capture process. The theoretical prediction is in good agreement with the experimental observation.

The Fig. 1(b) shows the distribution of $Q$ values for the double-electron capture process in $Ar^{2+}$-Ar collision system, in which the experimental data are represented by black scattered points, and the red curve corresponds to the reaction window calculated by MCBM. From the Fig. 1(b), it can be seen that the reaction channel of double electron capture can be expressed as: $Ar^{2+}$ ($3s^2 3p^4\ {}^3P, {}^1D, {}^1S$) + Ar (${}^1S$) → Ar (${}^1S$) + $Ar^{2+}$ ($3s^2 3p^4\ {}^3P, {}^1D, {}^1S$). The results of MCBM also show that the capture of the target electron to the ground state is dominant. Compared with the results of Kamber et al.[39], it is found that the peak position of the experimental results is shifted, which is mainly due to the fact that in Kamber et al.[39]'s experiment, there are only $Ar^{2+}$ ($3s^2 3p^4\ {}^3P$) ions in the ion beam, while in our experiment, there are metastable shells in addition to the recoil target ions produced by the ground state shells.

Shields and Moran[15] have calculated the potential energy curves for single and double electron capture reactions in $Ar^{2+}$ - Ar collision based on the Coulomb interaction and ion-induced dipole interaction proposed by Kamber et al.[40]. The internuclear distance corresponding to the intersection of the incident channel and the outgoing channel is called the capture internuclear distance $R$. According to the relationship between the reaction cross section and the capture internuclear distance in the classical model, $\sigma \propto \pi R^2$, the larger the capture interval $R$ is, the larger the reaction cross section will be. From the calculation results of Shields and Moran[15], it can be seen that the potential energy curves of the incident channel $Ar^{2+}$ ($3s^2 3p^4\ {}^3P, {}^1D, {}^1S$) + Ar (${}^1S$) and the outgoing channel $Ar^+$ ($3s3p^6\ {}^2S$) + $Ar^+$ ($3s^2 3p^5\ {}^2P^0$) intersect at about the internuclear distance of 13 a.u., and the potential energy curves of the outgoing channel $Ar^+$ ($3s^2 3p^5\ {}^2P^0$) + $Ar^+$ ($3s^2 3p^5$ ${}^2P^0$) is approximately at a nuclear spacing of 3.8 a.u. They intersect. This indicates that when the collision energy is low, the electron capture to the $Ar^+$ ($3s3p^6\ {}^2S$) state will be dominant, followed by the capture to the $Ar^+$ ($3s^2\ 3p^5\ {}^2p^0$) state, which is the same as our data. For the double electron capture process, Ar ($3s^2 3p^6$) is the main capture, and the theoretical prediction is in good agreement with the measured state population distribution.

By analyzing the scattering angular distribution, the dynamic mechanism of the electron capture process can be explored. The scattering angular distributions of single and double electron capture in $Ar^{2+}$-Ar collision system are shown in Fig. 2(a) and Fig. 2(b), respectively. The horizontal axis represents the scattering angle (in milliradians, mrad) and the vertical axis represents the collision count. The classical Rutherford model holds that the scattering angle is inversely proportional to the impact parameter $\theta \propto 1/b$. By analyzing the scattering angle distribution images of single electron capture, it can be observed that the collision reaction of capture to the ground state mainly occurs in a small scattering angle range. The angular distribution of single electron capture scattering calculated by MCBM is in good agreement with peak A in the experimental data. However, the angular distribution of peak B hardly changes with the scattering angle in the range of $\theta > 0.3$ mrad. This indicates that peak B may contain multiple reaction channels, and as the scattering angle increases, the impact parameter decreases, and the projectile ion prefers to be captured to a higher excited state. The angular distribution of double electron capture scattering calculated by MCBM is in good agreement with the experimental value. By comparing the Fig. 2(a) and Fig. 2(b), it can be found that the scattering angle corresponding to the double electron capture peak is larger than that corresponding to the single electron capture peak. Because MCBM assumes that electron capture occurs in an independent sequence and does not consider electron-electron interaction in the calculation of electron capture process, the model cannot provide accurate predictions for reactions involving multiple electrons, such as transfer target excitation, transfer projectile excitation, transfer ionization and so on. For these multi-electron reaction processes, more accurate theoretical models are needed for further study.

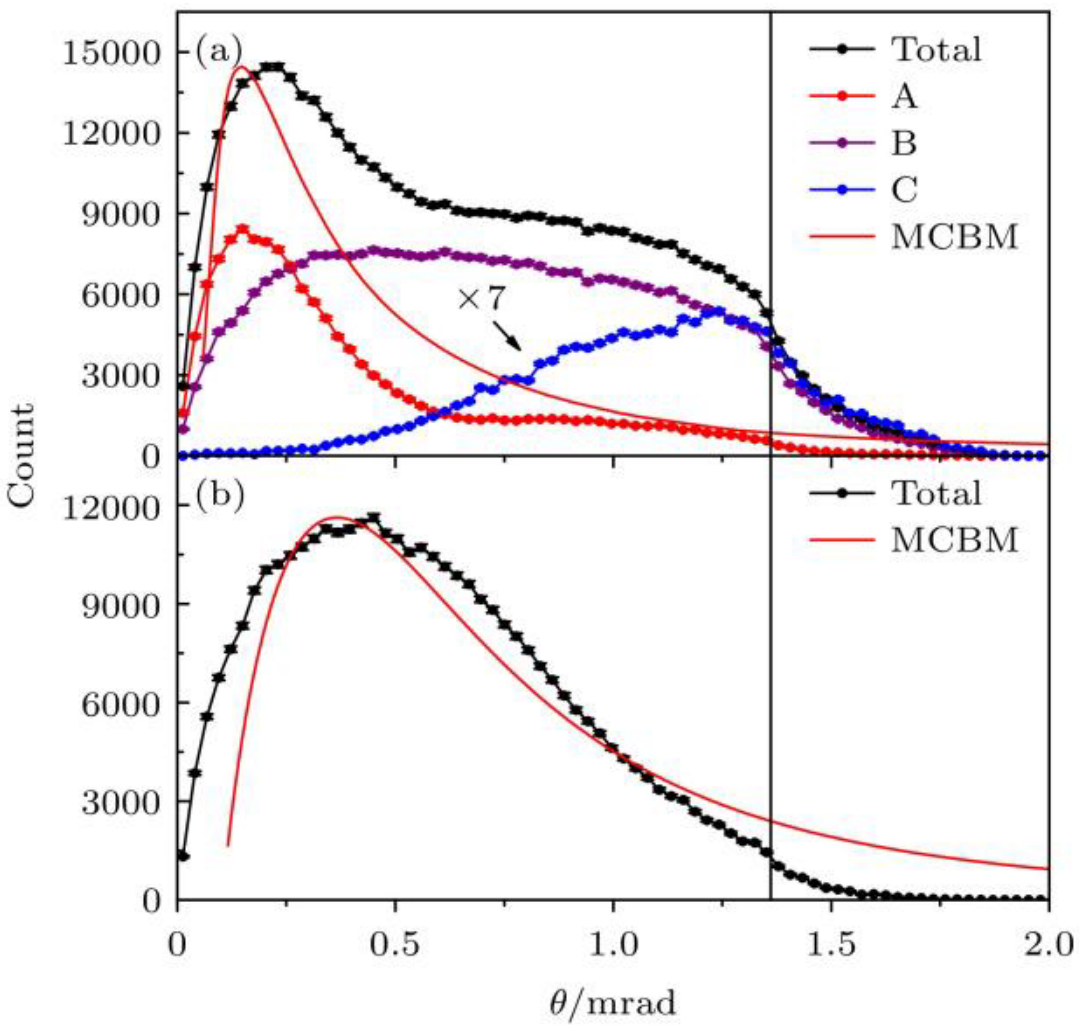

**Figure 2.** Scattering angle distributions for (a) single and (b) double electron capture in the collision process of $Ar^{2+}$-Ar at 40 keV energy. The red solid line represents the angular differential cross-section calculated by MCBM. The black vertical lines are auxiliary lines, and the region to the right of the vertical lines indicates the incompletely collected portion

## 3.2 $Ar^{2+}$-$N_2$ collision system

The Fig. 3(a) shows the $Q$ distribution of recoil ion $N_2^+$ in $Ar^{2+}$-$N_2$ collision system. The experimentally observed data (black data points) exhibit a bimodal structure, labeled peak A and peak B, respectively. It is worth noting that the single-electron capture channel of the $N_2$ target is similar to that of the Ar target, and the specific reaction mechanism is described as follows. A peak process: $Ar^{2+}$ ($3s^23p^4\,{}^3P,{}^1D,{}^1S$) + $N_2$ → $Ar^+$ ($3s^23p^5\,{}^2P^0$) +$N_2^+$, corresponding to $Q$ values of + 12.07 eV, + 13.70 eV and + 16.08 eV, respectively. B peak process: the main reaction channel is $Ar^{2+}$ ($3s^23p^4\,{}^3P,{}^1D,{}^1S$) + $N_2$ → $Ar^+$ ($3s3p^6\,{}^2S$) +$N_2^+$, corresponding to $Q$ values of – 1.41 eV, + 0.22 eV and + 2.60 eV, respectively, and in addition, $Ar^{2+}$ ($3s^23p^4\,{}^3P,{}^1D,{}^1S$) + $N_2$ → $Ar^+$($3s^23p^4nl$)+$N_2^+$. The process also makes certain contributions, with the corresponding $Q$ value range being -0.32 to 15.12 eV. Compared with the $Ar^{2+}$-Ar collision system, the reaction channels of $Ar^{2+}$-$N_2$ collision are more, which is mainly due to the existence of multiple different molecular configurations of $N_2^+$ ions[15]. It is worth noting that at 40 keV collision energy, the single-electron capture process $Q$ spectra of $Ar^{2+}$-$N_2$ and $Ar^{2+}$-Ar systems show significant similarity, which may be related to the close ionization energy of Ar atom (15. 76 eV) and $N_2$ molecule (15. 58 eV). Unlike the trimodal structure of Fig. 1(a), Fig. 3(a) exhibits only bimodal characteristics. This phenomenon can be attributed to the molecular dissociation process of $N_2^+$ ions in the energy range corresponding to the C peak in Fig. 1(a).

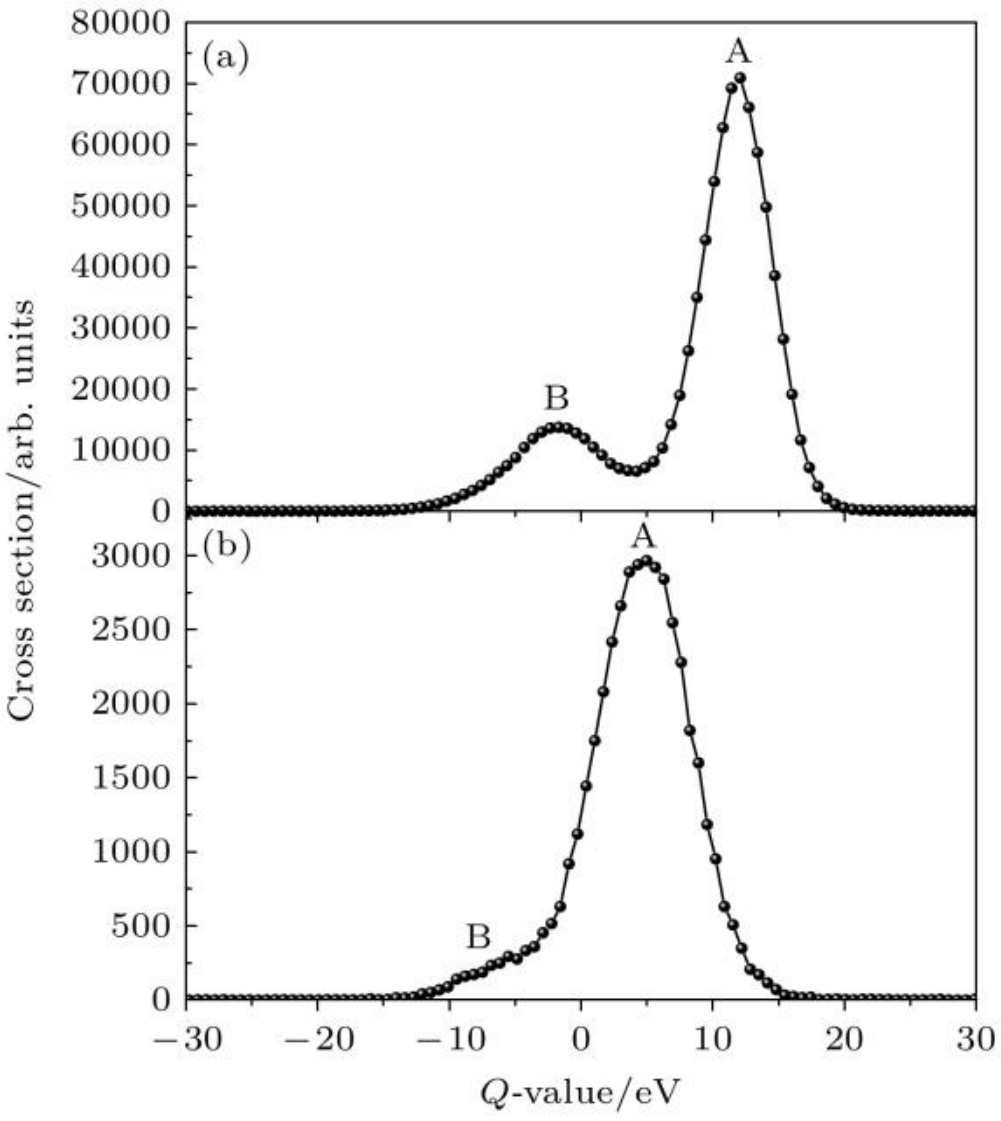

**Figure 3.** $Q$-value distributions for electron capture processes in $Ar^{2+}$-$N_2$ collisions at 40 keV: (a) Single electron capture; (b) double electron capture.

Figure 3(b) shows the Q-value distribution of recoil ions for double-electron capture in $Ar^{2+}$–$N_2$ collisions. The capture process of peak A in the figure is: $Ar^{2+}$ ($3s^2 3p^4\ {}^3P, {}^1D, {}^1S$) + $N_2$ → Ar ($^1S$) + $N_2^{2+}$, corresponding to $Q$ values of + 0.56 eV, + 2.19 eV and + 4.57 eV, respectively; the small peak B on the left of the figure corresponds to the process from the highly excited state to the ionization limit. The reaction process can be expressed as $Ar^{2+}$ ($3s^2 3\ p^4\ {}^3P, {}^1D, {}^1S$) + $N_2$ → Ar ($3s^2 3\ p^5 nl$) + $N_2^{2+}$, and the corresponding $Q$ values range from – 6.98 to – 15.20 eV.

The Fig. 4 shows the scattering angular distribution characteristics of single-electron capture and double-electron capture processes in $Ar^{2+}$-$N_2$ collision system. It can be seen from the experimental data that the maximum peak positions of the angular distribution for single-electron capture and double-electron capture processes are approximately equal. Compared with the Fig. 2, it can be observed that the peak value of the scattering angular distribution of the double-electron capture process of $Ar^{2+}$-Ar is obviously larger than that of the single-electron capture process. For the atomic target system, the number of electrons participating in the double-electron capture process is relatively more, which indicates that the effective impact parameter $b$ is smaller than that of the single-electron capture process. Based on the inverse relation between the scattering angle and the impact parameter $\theta \propto 1/b$, in the classical collision model, this model can reasonably explain that the peak value of the scattering angular distribution of the double-capture is greater than that of the single-capture in the $Ar^{2+}$-Ar collision process. In the molecular target system, when the distance between the projectile nucleus and the diatomic nucleus in the target molecule is close, the interaction effect between the nuclei will significantly affect the collision dynamics. The current theoretical framework cannot fully explain the complex interaction mechanism of this kind of ion-molecule collision system, and it is urgent to develop theoretical models including many-body interaction and molecular orbital characteristics to deepen the study of related mechanisms.

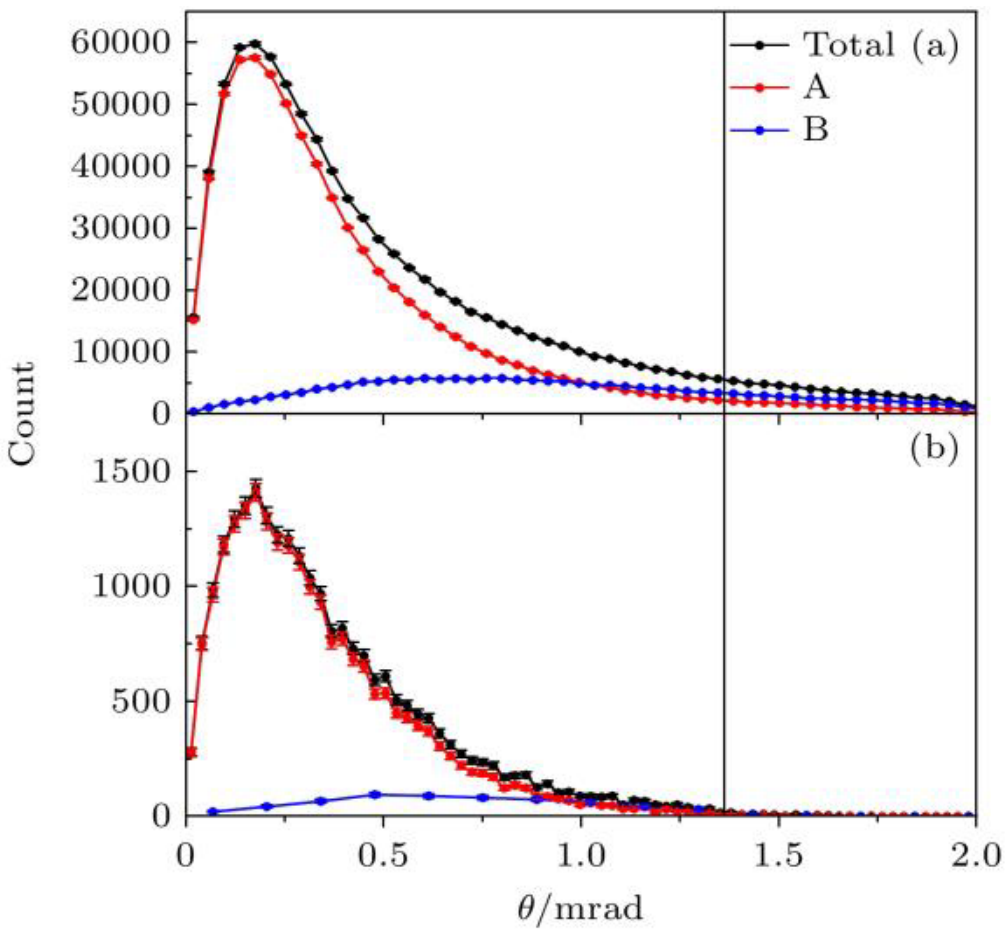

**Figure 4.** Scattering angle distributions for single and double electron capture in the collision process of $Ar^{2+}$-$N_2$ at 40 keV energy: (a) The single electron capture process; (b) the double electron capture process. The black vertical lines are auxiliary lines, and the region to the right of the vertical lines indicates the incompletely collected portion.

The Fig. 5 shows the distribution of the $Q$ -value of recoil ion $N_2^{2+}$ in $Ar^{2+}$-$N_2$ double electron capture process in different scattering angle range at 40 keV impact energy. By studying different scattering angle intervals, the distribution characteristics of the double-electron capture process in different spaces can be revealed. Only the characteristic peak (I) is observed Fig. 5(a), which corresponds to the ground-state double-electron capture reaction channel: $Ar^{2+}$ ($3s^2 3p^4\ {}^3P, {}^1D, {}^1S$) + $N_2$ → Ar (${}^1S$) +$N_2^{2+}$. The characteristic peak (II) can be clearly seen in the Fig. 5(b), and the corresponding double-electron capture reaction channel is $Ar^{2+}$ ($3s^2 3p^4\ {}^3P, {}^1D, {}^1S$) + $N_2$ → Ar ($3s^2 3p^5 nl$) + $N_2^{2+}$. In the scattering angle range of 0.2 - 0.6 mrad, the ground state capture channel is still dominant, but the secondary reaction channel (peak II) caused by Ar excited state capture can be clearly distinguished in the experimental spectrum. It is worth noting that the relative intensity of peak II in Fig. 5(c) increases significantly, which shows that the contribution of peak II increases with the increase of scattering angle (corresponding to the decrease of impact parameter $b$), while the intensity of peak I decreases significantly. This phenomenon indicates that for the $Ar^{2+}$-$N_2$ system, the probability of double electron capture to the ground state of Ar reaches its maximum in the small scattering angle region (large impact parameter), while the probability distribution of excited state capture (peak II) depends on the internuclear separation distance, and its contribution increases with the increase of scattering angle (decrease of internuclear distance). Generally speaking, under the condition of small scattering angle, only large impact parameter plays an important role in the electron capture process, because the avoided crossing point cannot be reached when the internuclear distance is large; When the scattering angle increases, the contribution of the smaller internuclear distance region gradually appears, which is consistent with the statistical distribution of the classical differential impact parameter[41].

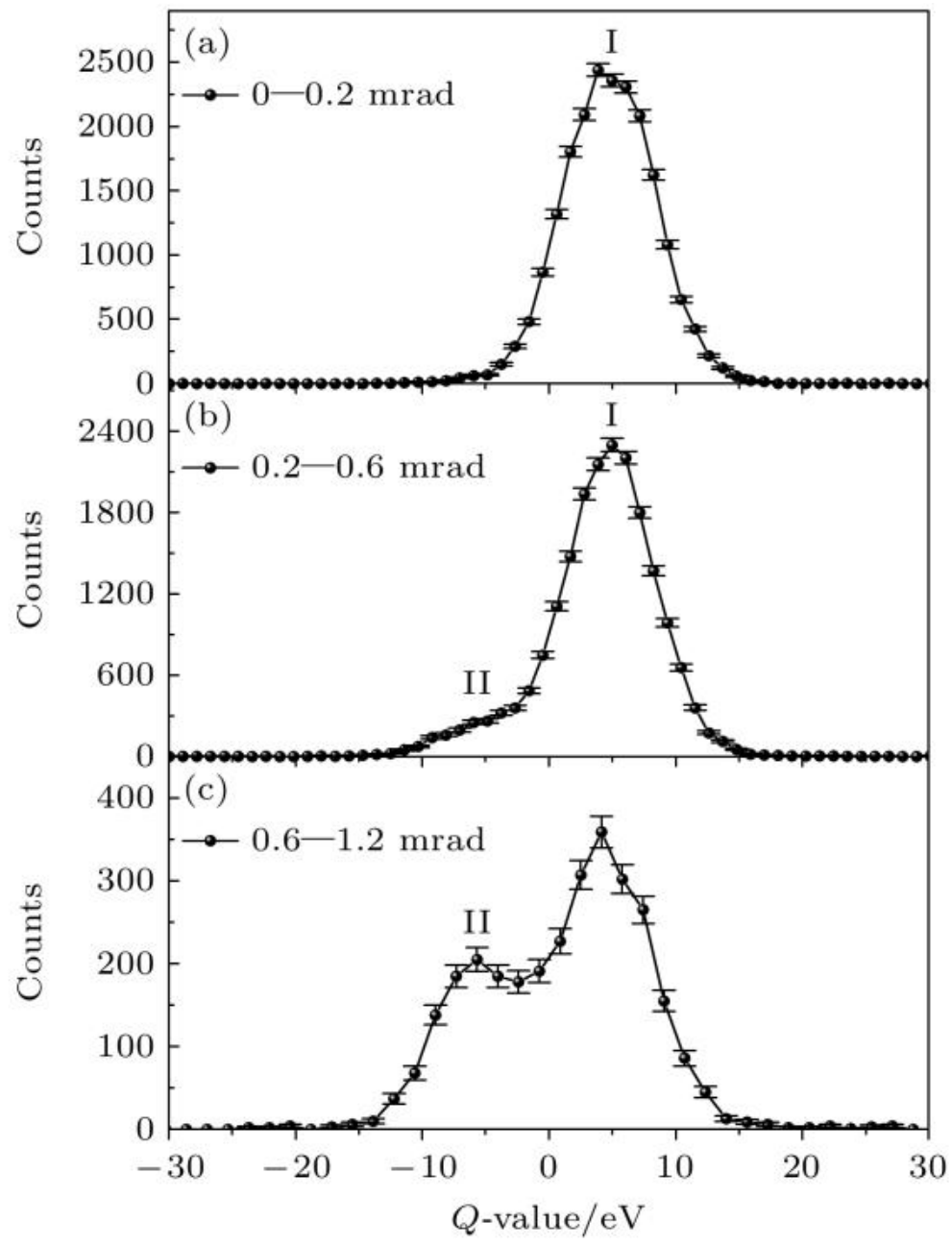

**Figure 5.** Spectrum of $Q$ values for double electron capture in the range of different scattering angles for the $Ar^{2+}$-$N_2$ collision process at 40 keV energy, where Process I: $Ar^{2+}$ ($3s^23p^4$ $^3P$, $^1D$, $^1S$) + $N_2 \rightarrow Ar(^1S) + N_2^{2+}$ and Process II: $Ar^{2+}$ ($3s^23p^4$ $^3P$, $^1D$,$^1S$)+$N_2 \rightarrow Ar(3s^23p^5 nl)$+ $N_2^{2+}$.

# 4. Conclusion

The electron capture dynamics in collisions of $Ar^{2+}$ ions with Ar/$N_2$ mixtures at 40 keV has been studied by using a reaction microscopic imaging spectrometer. The $Q$ spectra of state-resolved single and double electron capture processes for $Ar^{2+}$-Ar and $Ar^{2+}$ $N_2$ systems have been obtained by momentum imaging technique. The experimental results show that the populations of the single electron capture States of the $Ar^{2+}$-Ar and $Ar^{2+}$-$N_2$ systems are similar, but the contribution ratios of the states are different. In particular, an additional characteristic peak is observed in the $Q$ spectrum of the $Ar^{2+}$-Ar system, which corresponds to the process of the projectile ion capturing an electron from the 3s orbital of the target atom and exciting its 3s electron to the 3p orbital. However, in the $Ar^{2+}$-$N_2$ system, the characteristic peak does not appear because the $N_2^+$ ion is easy to dissociate after excitation. For the double electron capture process, the capture to the ground state is dominant in both systems, but a significant excited state population contribution is observed for the $Ar^{2+}$-$N_2$ system.

By comparing the scattering angles of the two collision systems, it is found that the higher the capture state of the product ion formed after the projectile captures an electron, the larger the corresponding scattering angle and the smaller the impact parameter. This may be due to the fact that at smaller impact parameters, the more electrons participate in the interaction, the more complex the capture reaction will be, and the probability of electrons being captured to high energy levels will be greater. The $Q$ spectrum of the double electron capture process of $Ar^{2+}$-$N_2$ in the range of 0 – 1.2 mrad shows that only the capture channel to the ground state will be filled at small angles. In addition, the analysis results also reveal an interesting phenomenon: the electron capture is dependent on the impact parameter, that is, with the increase of the angle (that is, the decrease of the impact parameter), the capture reaction tends to have a smaller $Q$ value, that is, the reaction tends to be more endothermic.